\documentclass[prb,reprint]{revtex4-2}

\usepackage[utf8]{inputenc}
\usepackage{amsmath}
\usepackage{amssymb}
\usepackage{graphicx}
\usepackage{color}
\usepackage{mathtools}
\usepackage{physics}
\usepackage{enumerate}
\usepackage[colorlinks=true, pdfstartview=FitV, linkcolor=black,citecolor=black, urlcolor=black]{hyperref}

\newcommand{\beq}[1]{\begin{equation}{\label{eq:#1}}}
\newcommand{\eeq}{\end{equation}}
\def\DD{{\mathchar'26\mkern-12mu d}}

\newcommand\pdt[3]{\left. \frac{\partial #1}{\partial #2} \right|_{\mathrlap{#3}}\hspace{3pt}}
\newcommand\pdtd[4]{\left. \frac{\partial #1}{\partial #2} \right|_{\mathrlap{#3}}\, \dd{#4}}
\newcommand\pdti[3]{\left[ \partial #1/\partial #2 \right]_{#3}}
\newcommand\pdtdi[4]{\left[ \partial #1/\partial #2 \right]_{#3}\, \dd{#4}}
\newcommand\pp{\mathcal{P}}
\newcommand\VV{\mathcal{V}}

\begin{document}
	
\author{Viktor Könye}
\affiliation{Institute for Theoretical Solid State Physics, IFW Dresden and W\"urzburg-Dresden Cluster of Excellence ct.qmat, Helmholtzstr. 20, 01069 Dresden, Germany}
\author{J{\'o}zsef Cserti}
\affiliation{Department of Physics of Complex Systems, E\"otv\"os Loránd University, H-1117 Budapest, P\'azm\'any P{\'e}ter s{\'e}t\'any 1/A, Hungary}

\date{\today}

\title{General formalism for calculating the thermal efficiency of thermodynamic cycles defined in a $p-V$ diagram}

\begin{abstract}
        We develop a general method for calculating the thermal efficiency of arbitrary thermodynamic cycles defined in the pressure-volume ($p-V$) diagram. To demonstrate how effective our approach is, we calculate the thermal efficiency of ideal gas engines for a few non-trivial cycles in the $p-V$ diagram, including a circular shape, a heart shape, a cycloid of Ceva, and a star-shaped curve. We determine the segments along the cycle where heat is absorbed or released from the heat engine. Our method can be applied to any gas model, and, as an example, we present the results for the van der Waals gas.
\end{abstract}

\maketitle

\section{Introduction}
\label{intro:sec}

Calculating the efficiency of thermodynamic cycles is a central issue in practical applications.
Moreover, when teaching, the basic concepts of thermodynamics can be introduced very efficiently through calculations of energy efficiency. 
In most textbook examples~\cite{Zemansky1997heat:book,Kubo_thermo_problems:book,Callen:450289}, the thermodynamic cycles in the pressure-volume ($p-V$) diagram include only four types of processes: \textit{isothermal} (constant temperature), \textit{isochoric} (constant volume), \textit{isobaric} (constant pressure), and \textit{adiabatic} (heat-insulated system).
Engines including such cycles, like the Carnot, Otto, Brayton (or Joule), Diesel, Stirling, Rankine, and Sargent cycles, are well known in the literature. 
To calculate the exchanged heat and work done for these systems, it is assumed that all processes are idealized; i.e., they are quasi-static and reversible, and there is no friction, turbulence, or heat loss. The common feature in the four above-mentioned types of idealized processes is that the exchanged heat and the work done can be calculated analytically for a gas characterized by its equation of state and its internal energy.

The calculation of the specific heat and of the thermal efficiency for \textit{unconventional}  thermodynamic processes (which deviate from the four special ones mentioned above) has been a frequently studied issue in the literature. 
For example, the specific heat has been calculated for gases in an arbitrary process in Ref.~\cite{doi:10.7227/IJMEE.29.3.5} and for the ideal gas along an elliptical $p-V$ cycle in Ref.~\cite{doi:10.1119/1.1495408}. 
Similarly, to calculate the thermal efficiency, the use of logarithmic plots of the $p-V$ diagram has been suggested in Ref.~\cite{doi:10.1119/1.4860656}.
The thermal efficiency has been calculated for cycles in the $p-V$ diagram involving diagonal processes with negative slope~\cite{doi:10.1119/1.17518,doi:10.1119/1.17944,doi:10.1119/1.18414,Arenzon_2018,Pacheco2021}, parabolic processes~\cite{Arenzon_2018},
circular processes~\cite{Marcella_2000}, an unconventional lobe~\cite{Chen_2006}, and an alternative thermodynamic cycle for the Stirling machine~\cite{doi:10.1119/1.5007063}.

Extending the calculation of the thermal efficiency to general cycles then is a natural development. 
Indeed, for unconventional processes, the calculation of the efficiency is not obvious at all  
since it is rather difficult to calculate the absorbed and released heat during the cycle. 
In principle, one needs to locate the segments of the cycle where heat is absorbed or released by the fluid. These segments are separated by points at which the adiabatic curves are tangent to the curve representing the given cycle~\cite{Pacheco2021}.
In most cases, such calculations cannot be performed analytically. 
This problem can be complicated even in the simple case of a circular process~\cite{Marcella_2000}.

\begin{figure}[ !ht]
	\centering
	\includegraphics[width=8.6cm]{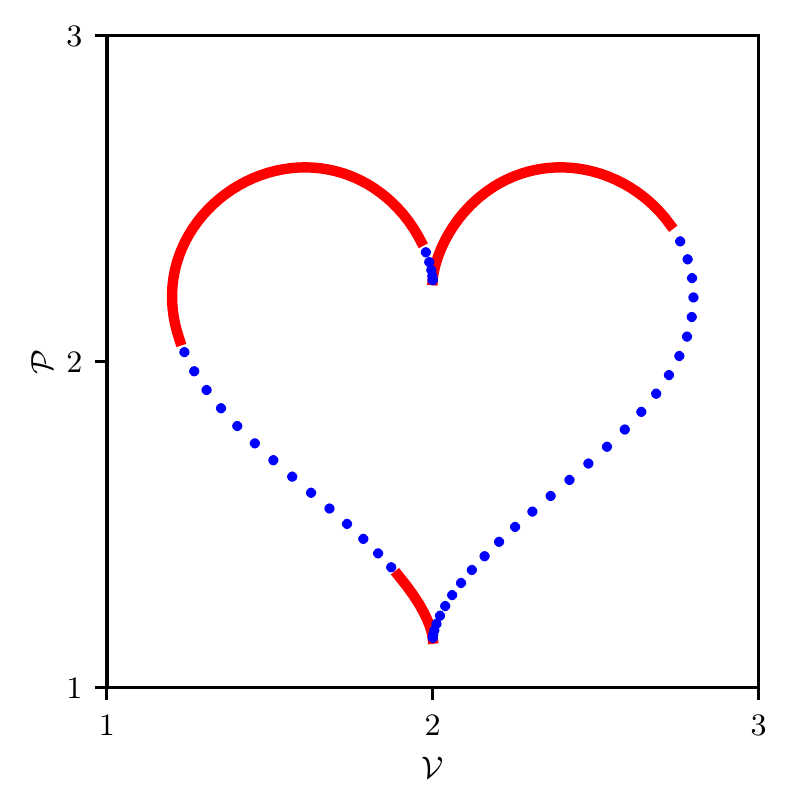}
	\caption{Heart-shaped curve in the $p-V$ diagram. 
		The parametric curve is given in the text (Eq.~(\ref{heart:eq})). 
		The red lines and the blue dotted lines correspond to segments where heat is absorbed and released 
		by the system, respectively.
		\label{fig:heart}}
\end{figure}

To avoid this problem, in this work, we present a general formalism for calculating the thermal efficiency for arbitrary cycles parametrized in the $p-V$ diagram. Figure~\ref{fig:heart} shows such an example in which the cycle is a heart-shaped curve.
Although such a heat engine is probably not realistic, our general approach could be useful in practice and may provide a didactically useful example for teaching the calculation of thermal efficiency as well.

Our formalism is quite versatile; it is valid for all gases provided their equation of state and their internal energy are known. 
We shall calculate the thermal efficiency both for ideal gas and for the van der Waals gas.   
From the calculation we shall identify the segments of the $p-V$ diagram where heat is absorbed or released from the heat engine.

The paper is organized as follows. 
In Sec.~\ref{gen_formlism:sec}, we present our general formalism to obtain the thermodynamic efficiency.
In  Sec.~\ref{application:sec}, we demonstrate the universality of our method by calculating the thermal efficiency of ideal gas engines for a few non-trivial processes defined in the $p-V$ diagram. 
In  Sec.~\ref{VdW_eta:sec}, we extend our formalism 
to the van der Waals gas. 
Finally, in Sec.~\ref{conclusions:sec}, we summarize our main conclusions. 

\section{General formalism}
\label{gen_formlism:sec}

First, we consider an arbitrary gas, assuming that its equation of state and its internal energy $U(T,V,n)$ are known. Here, $T$ is the temperature and $n$ is the number of moles. 
From the differential form of the first law of thermodynamics, 
the exchanged heat for an infinitesimal quasi-static process depending only on the thermodynamic coordinates $T$ and $V$ is given by
\begin{equation}
\label{DQ:eq}
\DD Q = \dd{U} + p\dd{V} = \pdtd{U}{T}{V}{T}
+\left( \pdt{U}{V}{T}+p\right) \dd{V}.
\end{equation}
The notation $\DD$ signals that $\DD Q$ is not an exact differential of the state variables.
The partial derivative of the internal energy $U$ in the first term  is the heat capacity at constant volume: $C_{V}=\pdti{U}{T}{V}$ (in general, it is known for a given gas),  
while the second term can be rewritten using the relation
\begin{equation}
\label{UderV:eq}
\pdt{U}{V}{T} +p = T\,\pdt{S}{V}{T}  = T \pdt{p}{T}{V} ,  
\end{equation}
which can be derived from the combined first and second laws of thermodynamics 
$\dd{U} = T \dd{S} - p\dd{V}$, where $S$ is the entropy, and the  Maxwell relation 
$\pdti{S}{V}{T} = \pdti{p}{T}{V}$. 
From the equation of state we may write 
$\dd{T}=\pdtdi{T}{p}{V}{p} + \pdtdi{T}{V}{p}{V}$,  and then Eq.~(\ref{DQ:eq}) can be rewritten as
\begin{align}
\label{DQ_1:eq}
\DD Q &= C_{V}\, \pdtd{T}{p}{V}{p} 
+\left(C_V \, \pdt{T}{V}{p} + T \pdt{p}{T}{V}\right)\, \dd{V} .
\end{align}
One can see that the exchanged heat $\DD Q $ can be obtained by knowing only the equation of state and the heat capacity at constant volume $C_V$.
Note that Eq.~\eqref{DQ_1:eq} is valid for all gases, and that the temperature dependence of the three partial derivatives can be eliminated using the equation of state, so that they depend only on $p$ and $V$.
Equation~(\ref{DQ_1:eq}) is one of the central results in this work and a starting point for further calculations.  

Let us now parameterize the cycle in the $p-V$ diagram.
For a general cycle, we define the dimensionless pressure $\pp(\varphi)$ and volume $\VV(\varphi)$ as a function of a parameter $\varphi$ in the following way:
\begin{align}
\label{eq:poVo}
p(\varphi) &= p_0\, \pp(\varphi), \quad \mathrm{and} 
\quad V(\varphi) = V_0\, \VV(\varphi),
\end{align} 
where $p_0$ and $V_0$ are constants characterizing the dimensions of $p$ and $V$, and the parameter $\varphi$ is a real number varying in a closed interval $\left[\varphi_i,\varphi_f\right]$. 
For a closed cycle, 
$p(\varphi_i)=p(\varphi_f)$ and $~V(\varphi_i)=V(\varphi_f)$. 
In this work, we consider only heat engines in which the cycle runs clockwise in the $p-V$ diagram, i.e., the point 
$\left( \VV(\varphi),\pp(\varphi) \right)$ on the curve of the cycle 
moves in the clockwise direction as $\varphi$ increases.  
Note that a similar approach has been used in Ref.~\cite{Chen_2006} for an unconventional lobe in the $p-V$ diagram with an ideal gas and Ref.~\cite{doi:10.1119/1.5007063} for a Stirling engine where the volume of the gas is parameterized by the camshaft angle $\theta$.
However, the formalism presented below is valid for arbitrary cycles and gases.

Since $\dd{p} = p_0(\dd{\mathcal{P}}(\varphi) / \dd{\varphi}) \dd{\varphi}$ and 
$\dd{V} = V_0(\dd{\mathcal{V}}(\varphi) /\dd{\varphi}) \dd{\varphi}$, the exchanged heat in Eq.~(\ref{DQ_1:eq}) along the curve is given by 
\begin{subequations}
\label{eq:heatchange} 
  \begin{align}
\DD Q &= p_0\, V_0\, q(\varphi)\dd{\varphi}, \quad \mathrm{where}\\
\nonumber
q(\varphi) &=  \frac{C_V}{V_0}\, \pdt{T}{p}{V}\, \frac{\dd{\mathcal{P}}(\varphi)}{\dd{\varphi}}+
\label{eq:q-fi} 
         \\&+\left(\frac{C_V}{p_0}\,   \pdt{T}{V}{p}
+\frac{T}{p_0}\,  \pdt{p}{T}{V}\right)\, \frac{\dd{\mathcal{V}}(\varphi)}{\dd{\varphi}} . 
\end{align} 
\end{subequations}
Note that to obtain the dimensionless heat function $q(\varphi)$, 
it is enough to know the equation of state of the gas, the heat capacity $C_V$ and the parametrized curve of the cycle in the $p-V$ diagram. 
When $q(\varphi)>0$ the system absorbs heat, and when $q(\varphi)<0$ heat is released from the system (according to the usual sign convention for heat transfers and work done). 
Note that a similar heat function has been introduced by Plischke and Bergersen in Problem 1.5 (see their book and the solution in Refs.~\cite{Plischke_book_1994,Plischke_book_1994_sulution}, respectively).

It is convenient to calculate the absorbed and released heat from the function $q(\varphi)$ by introducing two further functions:
\beq{q_pm:eq}
q_\pm =\frac{1}{2}\left(|q|\pm q\right).
\eeq
Using this definition, if $q>0$ then $q_+ =q$ and $q_-=0$, while for $q<0$ we have $q_+ =0$ and $q_-= -q>0$. 
In a closed cycle, it follows from the first law of thermodynamics  
that $Q_+ - Q_- +W =0$, where  $Q_+$ and $Q_-$  (both are positive numbers) are the total heat absorbed and released from the system, respectively, and $-W$ is the work done by the engine (for clockwise direction in the $p-V$ diagram, $W<0$).  
Using our heat functions $q_\pm(\varphi)$ we have 
\begin{subequations}
	\begin{align}
	\label{DQ_int:eq}
	Q_\pm &=p_0\,V_0\, \int\limits_{\varphi_i}^{\varphi_f} q_\pm(\varphi)\dd{\varphi}, \\
	|W| &=  Q_+ - Q_- = p_0\,V_0\, \int\limits_{\varphi_i}^{\varphi_f} q(\varphi)\dd{\varphi}. 
	\end{align}
\end{subequations}
Finally, the thermal efficiency of the heat engine can be written as 
\begin{equation}
\label{etadef:eq}
\eta = \frac{|W|}{Q_{+}} =  \dfrac{\int\limits_{\varphi_i}^{\varphi_f} q(\varphi)\dd{\varphi}}{\int\limits_{\varphi_i}^{\varphi_f} q_+(\varphi)\dd{\varphi}} =
2\,{\left(1+\dfrac{\int\limits_{\varphi_i}^{\varphi_f} |q(\varphi)|\dd{\varphi}}{\int\limits_{\varphi_i}^{\varphi_f} q(\varphi)\dd{\varphi}} \right)}^{-1}  .
\end{equation}
Equations~(\ref{eq:heatchange}) and~(\ref{etadef:eq}) are our main results in this work. 

In principle, to find the parts of the cycle where the heat is absorbed or released by the system, one needs to calculate the points (\textit{adiabatic points}) where the curves corresponding to adiabatic processes are tangent to the curve of the cycle~\cite{Chen_2006}. 
However, even for the simple circle-shaped cycle, it is a rather difficult task to determine algebraically or numerically these adiabatic points along 
the cycle (see, e.g., Ref.~\cite{Marcella_2000}).  
The adiabatic points are the zeros of the heat function $q(\varphi)$ given by Eq.~(\ref{eq:heatchange}).
But determining zeros is not an easy task either.

The advantage of our general formalism is that, knowing the heat function $q(\varphi)$ given by Eq.~(\ref{eq:q-fi}), the amount of absorbed and released heat can be easily calculated from 
Eq.~(\ref{DQ_int:eq}), \textit{without} knowing the location of the adiabatic points. 
Moreover, the sign of the function $q(\varphi)$ allows us to find numerically where the heat is absorbed or released from the heat engine.

As we shall see below, often the integrals in Eq.~(\ref{etadef:eq}) can only be evaluated numerically. 
In these cases, we also calculate the work done directly as 
\begin{align}
    W &= - \int\limits_{\varphi_i}^{\varphi_f}\, 
    p(\varphi)\dd{V(\varphi)}= 
    - p_0 V_0 \, \int\limits_{\varphi_i}^{\varphi_f}\, 
    \mathcal{P}(\varphi)\,
    \frac{\dd{\mathcal{V}}(\varphi)}{\dd{\varphi}}\, 
    \dd \varphi ,
\end{align} 
and the numerical precision can be controlled by checking whether the first law of thermodynamics is satisfied.
To compute the thermal efficiency from our formalism it is necessary to know the equation of state, the heat capacity and the path on the $p-V$ diagram. In practical applications these are not necessarily given analytically, but as discrete points. In these cases, the thermal efficiency can be approximated using standard numerical methods to compute the derivatives and integrals in Eqs.~(\ref{eq:heatchange}) and~(\ref{etadef:eq}).

Our method can also be used to calculate the entropy along the cycle 
in the $p-V$ diagram. 
Using Eq.~(\ref{eq:heatchange}) the change of entropy can be expressed in terms of $q$ as $\dd{S}= \DD Q/T = p_0\, V_0\, q(\varphi)\dd{\varphi}/T(\varphi) $. 

\section{Applications for ideal gas}
\label{application:sec}

In this section, we apply our general formalism to find the thermal efficiency for an ideal gas for which the equation of state and the internal energy are $pV = nRT$ and $U(T,V,n) = C_V T$ respectively,  
where $R$ is the molar gas constant.
We assume that the heat capacity $C_V$  is known for the given gas. 

Then, in Eq.~(\ref{eq:heatchange}) the partial derivatives can be easily  calculated from the equation of state, and the heat function $q(\varphi)$ becomes  
\begin{equation}
\label{DQ_4:eq}
q(\varphi) = \frac{1}{\gamma-1}\left[
\gamma \, \pp(\varphi)\, \frac{\dd{\mathcal{V}}(\varphi)}{\dd{\varphi}} 
+ \VV(\varphi)\, \frac{\dd{\mathcal{P}}(\varphi)}{\dd{\varphi}} \right],
\end{equation}
where we introduce the ratio of the heat capacities $\gamma=C_p/C_V$ (here $C_p$ is the heat capacity at constant pressure). 
In the derivation, we made use of Mayer's relation $C_p-C_V= nR$. 

From Eq.~(\ref{DQ_4:eq}) we can easily derive the heat function for the standard processes, i.e. the isobaric, isochoric, isothermal, and adiabatic processes, as follows:

\begin{enumerate}[(i)] 

\item for isobaric process (at dimensionless pressure $\pp_0$ between dimensionless volumes $\VV_1$ and $\VV_2$), 
$\VV(\varphi) = \VV_2\varphi+\VV_1(1-\varphi), \pp(\varphi) = \pp_0 $, $\varphi\in[0,1]$, the heat function is \\$q_p(\varphi) = \frac{\gamma}{\gamma-1}\pp_0(\VV_2-\VV_1)$,
\item for isochoric process (at dimensionless volume $\VV_0$ between dimensionless pressures $\pp_1$ and $\pp_2$), 
$\VV(\varphi) = \VV_0, \pp(\varphi) = \pp_1\varphi+\pp_2(1-\varphi) $, $\varphi\in[0,1]$, the heat function is \\$q_V(\varphi) = \frac{1}{\gamma-1}\VV_0(\pp_2-\pp_1)$,
\item for isothermal process (at temperature $T$ between dimensionless volumes $\VV_1$ and $\VV_2$),
$\VV(\varphi) = \varphi$, $\pp(\varphi) = nRT/(p_0V_0 \,\varphi)$, $\varphi\in[\VV_1,\VV_2]$, the heat function is $q_T(\varphi) = nRT/(p_0V_0 \,\varphi)$, which means that $\DD Q = p dV$ as we expect from the first law of thermodynamics, 
\item for adiabatic process from the parameterization $\VV (\varphi)= \varphi$, $\pp (t) = c/ \varphi^{\gamma}$
 (here $c$ is a constant), $\varphi\in[\VV_1,\VV_2]$, it follows that $q(\varphi) =0$ 
for all $\varphi$, as it should be. 
\end{enumerate}

From the heat function $q(\varphi)$ we can get the heat transfer $Q$ which agrees with the well-known textbook results, and derive the thermal efficiency of all well-known cycles mentioned in the introduction.

\subsection{Thermal efficiency of an ellipse in the $p-V$ diagram}
\label{id_gaz:sec}

\begin{figure}[!ht] 
	\centering
	\includegraphics[width=8.6cm]{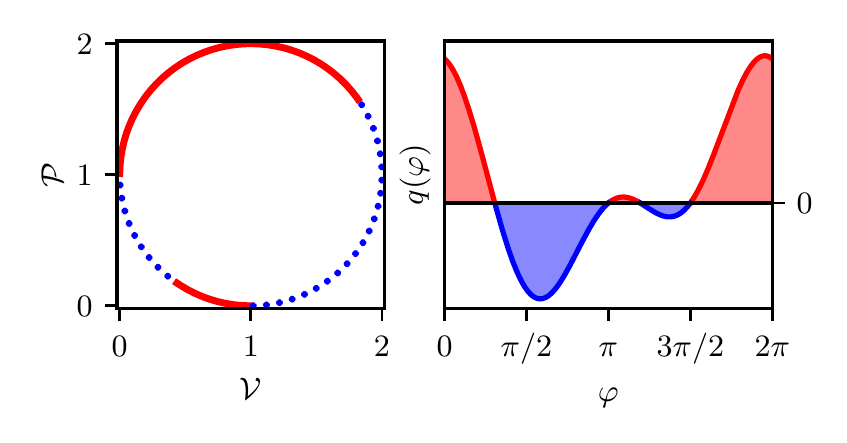}
        \caption{Circle-shaped cycle in the $p-V$ diagram given by Eq.~(\ref{circle:eq}) with $a=b=1$ (left panel) and the heat function $q(\varphi)$ (right panel). Solid red lines are used where $q(\varphi)>0$ and dotted blue lines where $q(\varphi)<0$.
                \label{fig:circle}
	}
\end{figure}

As a first example for applications, we consider a cycle described by an ellipse with axes parallel to the axes in the $p-V$ diagram (for a circular shape see the left panel in Fig.~\ref{fig:circle}).
This cycle can be parametrized as 
\begin{align}
\label{circle:eq}
\pp &= p/p_0 = 1+a\cos(\varphi)\quad \mathrm{and} \quad 
\VV = V/V_0 = 1+b\sin(\varphi),
\end{align}
where  $0\leq a,b\leq 1$.
Note that a similar parametrization has been used in  Ref.~\cite{doi:10.7227/IJMEE.29.3.5}.
Then, from Eq.~(\ref{DQ_4:eq}) we find 
\begin{align}
\label{q_ellipse_1:eq}
q(\varphi) &= \frac{ab}{\gamma-1}\left[\gamma 
\left(\frac{1}{a}+ \cos\varphi \right)\cos\varphi - \left(\frac{1}{b}+\sin\varphi \right)\sin\varphi\right].
\end{align}
Now, it is clear that the thermal efficiency given by Eq.~(\ref{etadef:eq}) depends only on $a$, $b$ and $\gamma$. 
In the case of $a=b=1$, the circular curve touches 
the $p$ and $V$-axes, as shown in Fig.~\ref{fig:circle} for a mono-atomic gas for which $\gamma = 5/3$. 

Although in general the zeros of $q(\varphi)$ cannot be calculated analytically, for $a=b=1$  these zeros can be determined:
\begin{subequations}
\begin{align}
\varphi_1 &= \arctan(\frac{\gamma-1+\sqrt{\gamma^2+6\gamma+1}}{1-\gamma+\sqrt{\gamma^2+6\gamma+1}}), \quad \varphi_2 = \pi,  \\
\varphi_3 &= \pi - \arctan(\frac{\gamma-1-\sqrt{\gamma^2+6\gamma+1}}{\gamma-1+\sqrt{\gamma^2+6\gamma+1}}), \quad \varphi_4 = \frac{3\pi}{2}. 
\end{align}
\end{subequations}
Thus, in this case, there are four sign changes 
in the heat function $q(\varphi)$, as can be seen in Fig.~\ref{fig:circle} 
and is predicted in Ref.~\cite{Liberto_2003}.
This case is quite unusual because, from our numerical calculation, we found that if the circle is a little bit further off from the $p$ and $V$-axes, then there are only two adiabatic points.    

Finally, the integrals in Eq. (\ref{etadef:eq})  can also be evaluated analytically:
\begin{align}
\int\limits_{0}^{2\pi} q(\varphi)\dd{\varphi} & =\pi \quad \mathrm{and} \quad  \int\limits_{0}^{2\pi} |q(\varphi)|\dd{\varphi}=
4\, \frac{\gamma^2+\gamma+1}{\gamma^2-1}, 
\end{align}
and thus for $a=b=1$, the thermal efficiency becomes a \textit{universal} 
value
\begin{align}
\label{eta_ellipse:eq}
\eta &=\frac{2}{1+\frac{4}{\pi}\, \frac{\gamma^2+\gamma+1}{\gamma^2-1}} .
\end{align}
Taking a mono-atomic gas with $\gamma=5/3$ we have $\eta= 8\pi/(49+4\pi)\approx0.4082$. 
Note that, as long as the ellipse touches the $p$ and $V$-axis, the thermal efficiency is a universal constant independent of $p_0$ and $V_0$.
Of course, this case is nonphysical since  the volume and the pressure cannot be zero during the cycle. 
Therefore, the above universal result can be interpreted only as a limiting case while $p\to0$ or $V\to 0$ at some points of the cycle.

Note that using Eqs.~(\ref{etadef:eq}) and~(\ref{DQ_4:eq}) 
we can easily find that $\eta=0.115$ which agrees with that given by Eq.~(18) in  Ref.~\cite{Marcella_2000} for a circular cycle with 
$\pp  = 1.5+ 0.5 \cos(\varphi)$, 
$\VV  = 1.5+0.5 \sin(\varphi)$, and $\gamma =1.4$. 
Another example in which our formalism can be tested is in Ref.~\cite{Liberto_2003} where it was shown that, for elliptical cycles 
the maximum of the thermal efficiency is obtained for $a=b=1$, and 
$\eta_{\mathrm{max}}< 0.3$. 
Indeed, from Eq.~(\ref{eta_ellipse:eq}), for $\gamma = 1.4$, we have
$\eta_{\mathrm{max}}=12 \pi /(109 + 6 \pi) \approx 0.295$. 

We would like to mention that in 2015 in the Rudolf Ortvay Competition in Physics, organized by Eötvös Loránd University problem 8 was related to the thermal efficiency of a circle-shaped cycle as the one discussed here~\cite{Ortvay_contest:site}. 
In this competition, the first author of this paper presented a solution to this problem which served as a basis for the present work.

\subsection{Thermal efficiency for other shapes in the $p-V$ diagram}

We now consider the heart-shaped cycle Fig.~\ref{fig:heart}. 
The curve can be parametrized as~\cite{MathWorld_heart:site}
\begin{subequations} 
	\label{heart:eq}
	\begin{align}
                \nonumber
	\label{heart_param:eq}
	\frac{p}{p_0} \equiv \mathcal{P} &= 
	2+ \frac{1}{20}\left[13\cos{(\varphi)}-5\cos{(2\varphi)}
-2\cos{(3\varphi)}-\right. \\ &\hspace{70pt}\left.-\cos{(4\varphi)}\right], \\ 
	\frac{V}{V_0}  \equiv \mathcal{V} &= 2+ \frac{4}{5} \sin^3(\varphi), 
	\end{align}
\end{subequations}
where $\varphi \in [0,2\pi]$. 

\begin{figure}[!ht] 
	\centering
	\includegraphics[width=8.6cm]{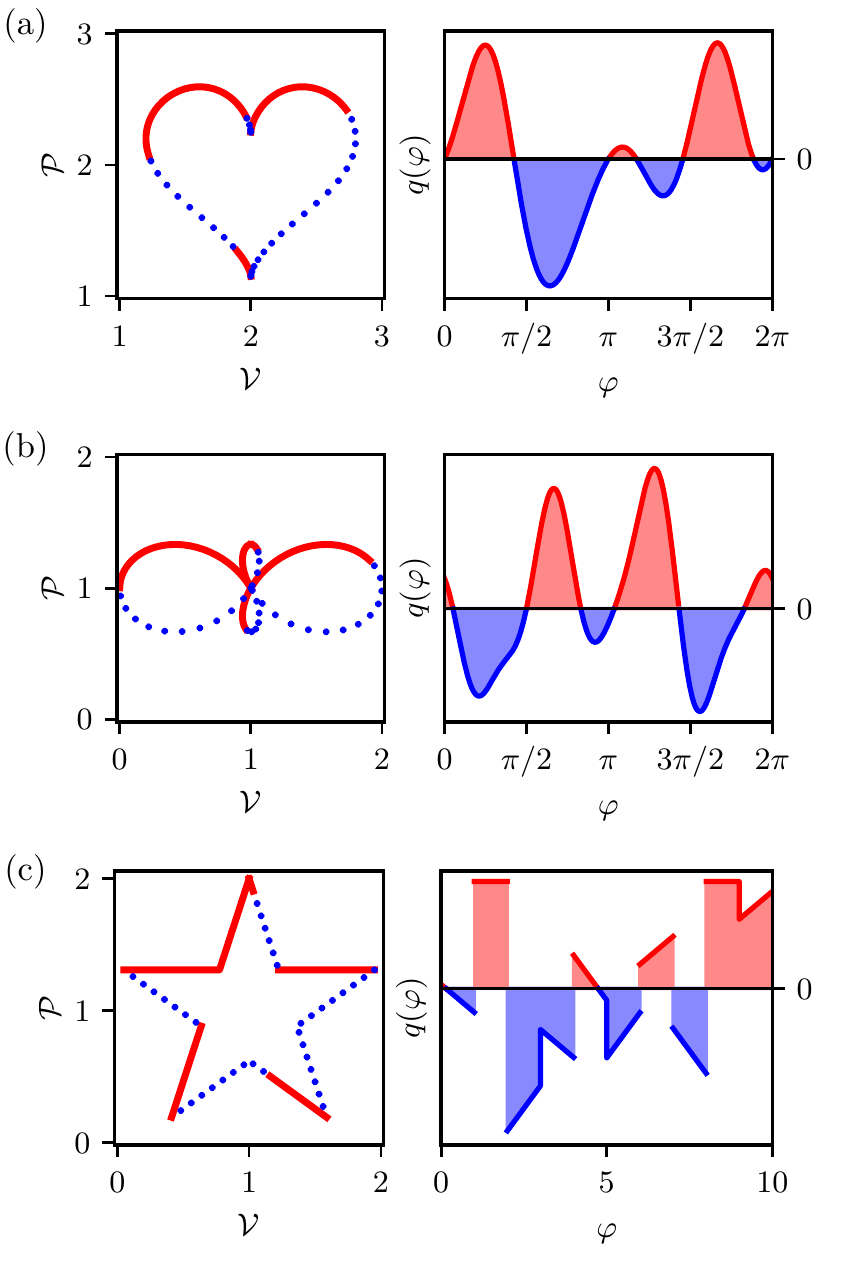}
        \caption{Thermal efficiency for different shapes in the $p-V$ diagram. (a) Heart-shape parametrized by Eq.~(\ref{heart:eq}), (b) Cycloid of Ceva parametrized by Eq.~(\ref{ceva:eq}), and (c) Star-shape. The left panels show the cycle in the $p-V$ diagram and the right panels show the corresponding heat function $q(\varphi)$. The coloring of the lines is the same as in Fig.~\ref{fig:circle}.
                \label{fig:shapes}
	}
\end{figure}

Again, from Eq.~(\ref{DQ_4:eq}) one can calculate the heat function $q(\varphi)$ analytically for this cycle and it is plotted in Fig.~\ref{fig:shapes}a (right panel).
However, the integrals in the expression for the thermal efficiency (Eq.~(\ref{etadef:eq}) can only be evaluated numerically. 
For a mono-atomic ideal gas ($\gamma= 5/3$) 
we find that $\eta = 0.1373$.

Our next non-trivial example is the cycloid of Ceva~\cite{Math_World_Ceva:site} which is parametrized by  
\begin{subequations}
    \label{ceva:eq}
	\begin{align}
	\frac{p}{p_0} &\equiv  \mathcal{P} =1+\frac{1}{3}\left[2\cos{(2\varphi)}-1\right]\cos{\varphi}, \\  
	\frac{V}{V_0} &\equiv \mathcal{V} =1+\frac{1}{3}\left[2\cos{(2\varphi)}-1 \right] \sin{\varphi}. 
	\end{align}
\end{subequations}
The shape of the curve and the corresponding heat function $q(\varphi)$ are shown in Fig.~\ref{fig:shapes}b.
For a mono-atomic ideal gas ($\gamma= 5/3$), 
we find numerically that $\eta = 0.1493$.

As shown by the previous two examples, it is quite clear that our method can be applied to cycles that can be described in polar coordinates.

Finally, we show that our method can be extended to cycles consisting of straight segments in the $p-V$ diagram. 
Any straight segment linking points $(V_A,p_A)$ and $(V_B,p_B)$ can be parametrized as 
\begin{subequations} 
\label{star_param:eq}
	\begin{align} 
	p(\varphi) &=p_A+ (p_B - p_A)\varphi , \\  
	V(\varphi) &=V_A+ (V_B - V_A)\varphi ,
	\end{align}
\end{subequations}
where the parameter is $\varphi\in\left[ 0,1\right] $.
Now, one can choose $V_0 = V_A$ and $p_0 = p_A$  in Eq.~(\ref{eq:poVo}).
Then the exchanged heat $\DD Q$ from Eq.~(\ref{eq:heatchange}) can be easily  calculated to obtain the thermal efficiency.  

Using the parametrization (\ref{star_param:eq}) and our general approach we can reproduce the results found in the literature. We obtain the same result as that of Problem 1.5 in the book by Plischke and Bergersen~\cite{Plischke_book_1994,Plischke_book_1994_sulution}.
Moreover, our calculation leads to the same thermal efficiency $\eta = 16/97 \approx 0.165$ for the cycle shown in Fig.~1 of Ref.~\cite{doi:10.1119/1.17518}.
Similarly, using our formalism, we easily verified the result shown in Fig.~4 of Ref.~\cite{Arenzon_2018}.

Finally, we use this type of parametrization to calculate the thermal efficiency of the star-shaped cycle shown in Fig.~\ref{fig:shapes}c.   
For a mono-atomic ideal gas ($\gamma= 5/3$), we find numerically that $\eta = 0.1453$.

\section{Thermal efficiency for the van der Waals gas}
\label{VdW_eta:sec}

Our method to calculate the thermal efficiency may be applied not only to the ideal gas but to more realistic ones such as the van der Waals gas. 
The equation of state proposed first by van der Waals takes into account particle interactions and the finite size of the particle. It is given by 
\begin{subequations}
	\label{VdW_stat_U:eq}
	\begin{align}
	\label{VdW_state:eq}
	\left(p+\frac{a n^2}{V^2}\right) \left(V-b n\right) &= nRT,\\
	\label{VdW_U:eq}
	U(T,V,n) &= C_V T - \frac{an^2}{V}, 
	\end{align}
\end{subequations}
where $a>0$ and $b>0$ are constants characterizing the interactions and the size of the particles of a given gas. 

Introducing the dimensionless variables $\hat{V}= V/V_c$,  $\hat{p}= p/p_c$, and $\hat{T}= T/T_c$ the equation of state (\ref{VdW_state:eq}) can be written in a universal form independent of the parameters of the gas: 
\begin{align}
\label{VdW_red:eq}
\left(\hat{p}+ \frac{3}{\hat{V^2}} \right) \left( 3\hat{V}-1\right) &= 8 \hat{T}, 
\end{align}
where $V_c  =  3 bn $, $p_c  = \frac{1}{27}\, \frac{a}{b^2}$ and 
$R T_c = \frac{8}{27}\,  \frac{a}{b}$ are the volume, pressure and the temperature at the critical point of the van der Waals gas.
This is the law of corresponding states.

If we now choose the previously defined parameters in Eq. (\ref{eq:poVo}) as $p_0=p_c$ and $V_0=V_c$ we get $\pp\equiv\hat{p}$ and $\VV\equiv\hat{V}$.
Now, using Eqs.~(\ref{eq:heatchange}) and~(\ref{VdW_stat_U:eq}) we find that the heat function is

\begin{align}
        \nonumber
        q(\varphi) = \frac{1}{\Gamma-1}&\left\{\left[
\Gamma \, \left(\pp+\frac{3}{\mathcal{V}^2}\right)-2\frac{3\VV-1}{\VV^3}\right]\, \frac{\dd{\mathcal{V}}}{\dd{\varphi}} 
                +\right.\\  &\hspace{50pt}+\left. \frac{3\VV-1}{3}\, \frac{\dd{\mathcal{P}}}{\dd{\varphi}} \right\},
\end{align}
where $\Gamma = 1+ \frac{nR}{C_v}$.
Note that for the ideal gas $\Gamma = \gamma = C_p/C_V$. 

\begin{figure}[!ht] 
	\centering
	\includegraphics[width=8.6cm]{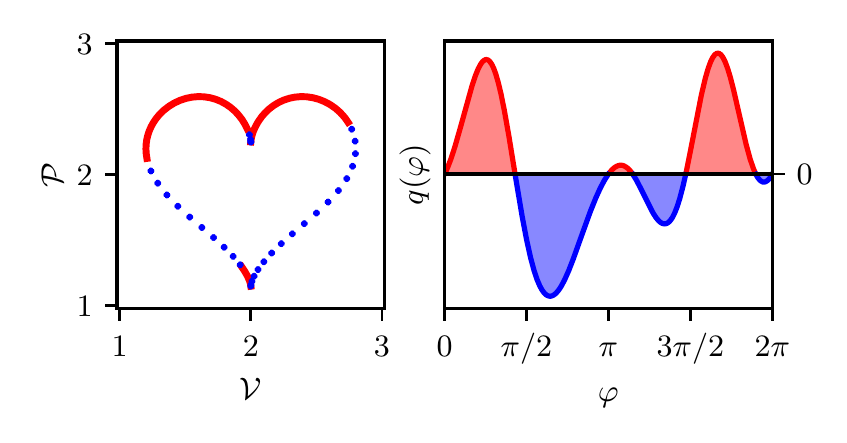}
	\caption{\label{heart_3:fig}The heart-shaped cycle using the van der Waals gas approximation.	The parameter $\Gamma=5/3$ is chosen based on the heat capacity of argon.}
\end{figure}

We now calculate the thermal efficiency and the heat function $q(\varphi)$ of the heart-shaped curve defined in Eq. (\ref{heart:eq}) using the van der Waals gas (see Fig. \ref{heart_3:fig}). The curve is chosen so that $T>T_c$ is always satisfied throughout the cycle. Using $\Gamma\approx5/3$ calculated from the heat capacity of argon, we get $\eta=0.1396$. As it is expected the difference between the monoatomic ideal gas and a more realistic model for argon is very small above the critical temperature.

\section{Conclusions}
\label{conclusions:sec}

We developed a general formalism to calculate the thermal efficiency of heat engines for arbitrary cycles defined in the $p-V$ diagram. Our method can be applied to all gases  provided their equation of state and their heat capacity are known. We derived a general and rather simple expression for the thermal efficiency.  
To demonstrate how versatile our formalism is, we calculated the thermal efficiency for a few  rather non-trivial cycles. 
 Finally, we derived the heat function for the van der Waals gas. A python code for calculating the thermal efficiency of the cycles presented in this work is available in the online supplemental information.

We believe that our general formalism provides a convenient and practical method to calculate the thermal efficiency of thermodynamic cycles defined on a $p-V$ diagram, without the need to locate the adiabatic points. Moreover, from a practical point of view, our method can also be applied to model more general cycles, different from the ones usually described in textbooks. We hope that our work can be useful for physics teachers and students to get more insight into the calculation of thermal efficiency.

\begin{acknowledgments}
        In memory of Gy. Radnai.	
	We would like to thank Gy.\ Dávid for helpful discussions.
	This work was supported by NKFIH within the Quantum Technology National Excellence Program (Project No. 2017-1.2.1-NKP-2017-00001) and 
	within the Quantum Information National Laboratory of Hungary,
	by the ELTE Institutional Excellence Program (TKP2020-IKA-05) financed by the Hungarian Ministry of Human Capacities, and Innovation Office (NKFIH) through Grant Nos. K134437.	
	
\end{acknowledgments}

\end{document}